\documentclass[a4paper,11pt]{article}
\usepackage{jinstpub}
\usepackage{lineno}
\usepackage{natbib}
\usepackage{booktabs}
\usepackage{multirow}
\usepackage{hyperref}
\usepackage{subcaption}
\usepackage{soul}

\title{Internal Calibration of the PandaX-II Detector with Radon Gaseous Sources}

\author        [a]        {Wenbo Ma,}
\author        [a]        {Abdusalam Abdukerim,}
\author        [a]        {Zihao Bo,}
\author        [a]        {Wei Chen,}
\author        [a,b]      {Xun Chen,}
\author        [c]        {Yunhua Chen,}
\author        [d]        {Chen Cheng,}
\author        [e]        {Xiangyi Cui,}
\author        [f]        {Yingjie Fan,}
\author        [g]        {Deqing Fang,}
\author        [g]        {Changbo Fu,}
\author        [h]        {Mengting Fu,}
\author        [i,j]      {Lisheng Geng,}
\author        [a]        {Karl Giboni,}
\author        [a]        {Linhui Gu,}
\author        [c]        {Xuyuan Guo,}
\author        [a,1]      {Ke Han,\note{Corresponding author}}
\author        [a]        {Changda He,}
\author        [c]        {Shengming He,}
\author        [a]        {Di Huang,}
\author        [c]        {Yan Huang,}
\author        [k]        {Yanlin Huang,}
\author        [a]        {Zhou Huang,}
\author        [l]        {Xiangdong Ji,}
\author        [m]        {Yonglin Ju,}
\author        [e]        {Shuaijie Li,}
\author        [m]        {Huaxuan Liu,}
\author        [a,b,e,2]  {Jianglai Liu,\note {Spokesperson}}
\author        [g,n]      {Yugang Ma,}
\author        [h]        {Yajun Mao,}
\author        [a,b]      {Yue Meng,}
\author        [a]        {Kaixiang Ni,}
\author        [c]        {Jinhua Ning,}
\author        [a]        {Xuyang Ning,}
\author        [o]        {Xiangxiang Ren,}
\author        [c]        {Changsong Shang,}
\author        [a]        {Lin Si,}
\author        [i]        {Guofang Shen,}
\author        [l]        {Andi Tan,}
\author        [o]        {Anqing Wang,}
\author        [n,p]      {Hongwei Wang,}
\author        [o]        {Meng Wang,}
\author        [n,q]      {Qiuhong Wang,}
\author        [h]        {Siguang Wang,}
\author        [d]        {Wei Wang,}
\author        [m]        {Xiuli Wang,}
\author        [a,b]      {Zhou Wang,}
\author        [d]        {Mengmeng Wu,}
\author        [c]        {Shiyong Wu,}
\author        [a]        {Weihao Wu,}
\author        [a]        {Jingkai Xia,}
\author        [l,r]      {Mengjiao Xiao,}
\author        [e]        {Pengwei Xie,}
\author        [a]        {Binbin Yan,}
\author        [a]        {Jijun Yang,}
\author        [a]        {Yong Yang,}
\author        [f]        {Chunxu Yu,}
\author        [o]        {Jumin Yuan,}
\author        [a]        {Ying Yuan,}
\author        [c]        {Jianfeng Yue,}
\author        [a]        {Xinning Zeng,}
\author        [l]        {Dan Zhang,}
\author        [a,b]      {Tao Zhang,}
\author        [a,b]      {Li Zhao,}
\author        [k]        {Qibin Zheng,}
\author        [c]        {Jifang Zhou,}
\author        [a]        {Ning Zhou,}
\author        [i,1]      {and Xiaopeng Zhou}
\collaboration {PandaX-II Collaboration}
\affiliation   [a]  {INPAC and School of Physics and Astronomy, Shanghai Jiao Tong University, MOE Key Lab for Particle Physics, Astrophysics and Cosmology, Shanghai Key Laboratory for Particle Physics and Cosmology, Shanghai 200240, China}
\affiliation   [b]  {Shanghai Jiao Tong University Sichuan Research Institute, Chengdu 610213, China}
\affiliation   [c]  {Yalong River Hydropower Development Company, Ltd., 288 Shuanglin Road, Chengdu 610051, China}
\affiliation   [d]  {School of Physics, Sun Yat-Sen University, Guangzhou 510275, China}
\affiliation   [e]  {Tsung-Dao Lee Institute, Shanghai 200240, China}
\affiliation   [f]  {School of Physics, Nankai University, Tianjin 300071, China}
\affiliation   [g]  {Key Laboratory of Nuclear Physics and Ion-beam Application (MOE), Institute of Modern Physics, Fudan University, Shanghai 200433, China}
\affiliation   [h]  {School of Physics, Peking University, Beijing 100871, China}
\affiliation   [i]  {School of Physics, Beihang University, Beijing 100191, China}
\affiliation   [j]  {International Research Center for Nuclei and Particles in the Cosmos \& Beijing Key Laboratory of Advanced Nuclear Materials and Physics, Beihang University, Beijing 100191, China}
\affiliation   [k]  {School of Medical Instrument and Food Engineering, University of Shanghai for Science and Technology, Shanghai 200093, China}
\affiliation   [l]  {Department of Physics, University of Maryland, College Park, Maryland 20742, USA}
\affiliation   [m]  {School of Mechanical Engineering, Shanghai Jiao Tong University, Shanghai 200240, China}
\affiliation   [n]  {Shanghai Institute of Applied Physics, Chinese Academy of Sciences, Shanghai 201800, China}
\affiliation   [o]  {School of Physics and Key Laboratory of Particle Physics and Particle Irradiation (MOE), Shandong University, Jinan 250100, China}
\affiliation   [p]  {Shanghai Advanced Research Institute, Chinese Academy of Sciences, Shanghai 201210, China}
\affiliation   [q]  {University of Chinese Academy of Sciences, Beijing 100049, China}
\affiliation   [r]  {Center for High Energy Physics, Peking University, Beijing 100871, China}

\emailAdd{ke.han@sjtu.edu.cn; zhou\_xp@buaa.edu.cn}

\abstract{
We have developed a low-energy electron recoil (ER) calibration method with $^{220}$Rn for the PandaX-II detector.
$^{220}$Rn, emanated from natural thorium compounds, was fed into the detector through the xenon purification system.
From 2017 to 2019, we performed three dedicated calibration campaigns with different radon sources.
We studied the detector response to $\alpha$, $\beta$, and $\gamma$ particles with focus on low energy ER events.
During the runs in 2017 and 2018, the amount of radioactivity of $^{222}$Rn were on the order of 1\% of that of $^{220}$Rn and thorium particulate contamination was negligible, especially in 2018.
We also measured the background contribution from $^{214}$Pb for the first time in PandaX-II with the help from a $^{222}$Rn injection.
Calibration strategy with $^{220}$Rn and $^{222}$Rn  will be implemented in the upcoming PandaX-4T experiment and can be useful for other xenon-based detectors as well.
}


\begin{document}

\maketitle
\section{Introduction}
While the existence of dark matter in the Universe is firmly established by cosmological observations~\cite{Planck}, the possible particle nature of dark matter is still unknown and widely studied~\cite{Schumann:2019eaa, JianglaiReview, Baudis:2012ig}.
Direct detection experiments examine the possible scattering of galactic dark matter particles on ordinary matter in detectors.
So far the most sensitive direct detection results, specifically for Weakly Interacting Massive Particles (WIMPs) at mass range around 100~$\rm{GeV/c^2}$, are given by dual-phase xenon Time Projection Chambers (TPC)~\cite{1-ton-year, 54-ton-day, LUX2016}.

A dual-phase xenon TPC excels in discriminating possible WIMP signals from background events with detector response.
A scattering event gives a prompt scintillation signal (usually called $S1$) and a delayed electroluminescence signal ($S2$) when drifted ionization electron enters a stronger electric field in the gas phase.
WIMP would most likely scatter with xenon nuclei and trigger a nuclear recoil (NR) signal.
On the contrary, electron recoil (ER) events are predominantly from $\beta$ and $\gamma$ particles interacting with electrons in xenon atoms.
NR and ER events differ from the relative amplitude of their corresponding $S2$ and $S1$ signals ($S2/S1$) and therefore ER backgrounds can be effectively rejected.

The PandaX-II detector was operated from 2015 to 2019 in China Jin-Ping underground Laboratory (CJPL-I)~\cite{CJPL} to search for WIMPs as well as other dark matter candidates and neutrinoless double $\beta$ decay~\cite{Ni:2019kms, spin-dependent, axion, SIDM, EFT, inelastic}.
  The PandaX-II detector had an active mass of 580~kg of xenon in a cylindrical field cage with a diameter of 646~mm and height of 600~mm, which corresponding to a maximum drift time of 360~$\mu$s~\cite{andiPhD}.
The top and bottom ends of the TPC were instrumented with 110 Hamamatsu-R11410 3-inch photomultiplier tubes (PMTs).
A possible WIMP signal or background events with energy larger than a few keV would trigger the data acquisition (DAQ) to record 1~ms-long waveform.
During physics data taking, the trigger rate was around 3~Hz.
Offline analysis identified $S1$ and $S2$ signals from the waveforms and the charge of  $S1$ and $S2$ signals were used to reconstruct the position and energy of each event with the help of different calibration sources.
We have calibrated the detector with external gamma and neutron sources as well as internal radioactive sources, which is the main topic of this paper.

Calibration with an internal $\beta$ source provides a uniformly distributed low-energy ER events to help determine a detector-specific ER band in the ($S1$, $S2/S1$) phase space.
Multiple types of sources, including CH${_3}$T~\cite{Akerib:2015wdi}, $^{228}$Th-based $^{220}$Rn~\cite{Aprile:2016pmc}, $^{37}$Ar~\cite{Boulton:2017hub}, $^{127}$Xe~\cite{Akerib:2017hph}, and $^{83\text{m}}$Kr~\cite{Akerib:2017eql}, have been developed by several collaborations for this purpose.
A few multi-ton scale dual-phase xenon TPCs are currently under construction~\cite{LZintro, XENONnTtalk, panda4simu} to keep searching for WIMP signals.
As the dimensions of those detectors get larger, it becomes more difficult for external sources to provide enough statistics at the center of the detectors in a reasonable time scale.
Internal calibrations with gaseous sources are becoming even more essential.

In this paper, we report the results from three internal calibration campaigns with radon sources at PandaX-II.
Three different sources were inserted into the xenon online purification loop and emanated $^{220}$Rn reached the TPC under the influence of gas flow.
$\beta$ decay of the daughter nucleus $^{212}$Pb gave low energy ER events that were uniformly distributed in the detector volume.
A large sample of $\alpha$ events from $^{220}$Rn and $^{216}$Po offered calibration at MeV scale, which is particularly important for future large scale multi-purpose TPCs.
We collected about ten thousand ER events in the dark matter search region of interest (ROI) from 0 to 20~keV$_{ee}$.
A data-driven ER distribution model in the ($S1$, $S2/S1$) phase space was constructed based on those events and used for dark matter analysis \cite{final-analysis, Zhou:2020bvf} .
The temporal event rate evolution and contaminations of radon injection were studied with high energy $\alpha$ events, low energy $\beta$ events, and $\beta-\alpha$ coincidence events.
We also presented a new method to precisely measure the $^{214}$Pb activity, which is the major background source for PandaX-II and other large scale xenon-based dark matter direct search  detectors.

\section{Radon Calibration Sources}

$^{220}$Rn sources made from natural thorium compounds were used in PandaX-II.
$^{220}$Rn is a noble gas isotope with a half-life of 55~s and part of $^{232}$Th decay chain.
$^{232}$Th has a natural isotopic abundance of 99.98\% with a long half-life of 1.4 billion years.
Possible chemical processing would break the secular equilibrium of the $^{232}$Th decay chain but keep its fast-decaying daughter $^{228}$Th, which decays to $^{224}$Ra and then $^{220}$Rn.
$^{220}$Rn gas emanates from the solid and may be injected into the detector for calibration.
$^{220}$Rn and its daughters decay to insignificant level within a couple of days after injection and thus a good choice for internal calibration.
However, the main concerns for natural thorium sources are contaminations of $^{222}$Rn and particulate thorium compounds.
$^{222}$Rn originates from $^{230}$Th and takes weeks to decay away, and the latter introduces permanent backgrounds to the detector.
Decay data of  $^{220}$Rn, $^{222}$Rn gas, and relevant progenies are shown in Table~\ref{tab:decay-chain}.

\begin{table}[tb]
\centering
\begin{tabular}{p{0.3\textwidth}cccccccc}
\toprule
Isotope             & $^{220}$Rn & $^{216}$Po & $^{212}$Pb & $^{212}$Bi       & $^{212}$Po \\
Half-life                    & 55~s       & 0.14~s     & 10.6~h     & 61~m             & 299~ns     \\
Decay mode                   & $\alpha$   & $\alpha$   & $\beta$    & $\beta$ (64.1\%) & $\alpha$   \\
E ($\alpha$) or Q-value [MeV] & 6.288      & 6.778      & 0.574      & 2.254            & 8.784      \\
\toprule
Isotope             & $^{222}$Rn & $^{218}$Po & $^{214}$Pb & $^{214}$Bi        & $^{214}$Po \\
Half-life                    & 3.8~d      & 3.1~m      & 26.8~m     & 19.9~m            & 164~$\mu$s \\
Decay mode                   & $\alpha$   & $\alpha$   & $\beta$    & $\beta$ (99.98\%) & $\alpha$   \\
E ($\alpha$) or Q-value [MeV] & 5.490      & 6.002      & 1.024      & 3.272             & 7.687      \\
\bottomrule
\end{tabular}
\caption{
	Decay data of $^{220}$Rn, $^{222}$Rn, and relevant progenies.
	$^{212}$Po decays to the stable $^{208}$Pb.
	The daughter of $^{214}$Po, $^{210}$Pb, has a half-life of 22.3~y and thus contribute negligible amount of events during calibrations.
	$\alpha$ decay branch of $^{212}$Bi and $^{214}$Bi is not used in this work and not shown.
}
\label{tab:decay-chain}
\end{table}

The $^{220}$Rn emanation rate varies for different thorium sources, depending on the mechanical structure, gas flow rate, and arrangement inside the gas loop.
The relative activities of emanated $^{220}$Rn and $^{228}$Th is measured as an indicator of emanation efficiency.
We arranged the sources in a source chamber and flushed it with boil-off nitrogen gas at a flow rate of 2 standard liter per minute (SLPM) and activities of $^{220}$Rn is measured by a commercial RAD7 radon detector~\cite{rad7} downstream.
The radioactivity of $^{228}$Th is measured by its characteristic $\gamma$ peaks with a High Purity Germanium (HPGe) detector~\cite{Xuming}.
We measured activities of emanated $^{222}$Rn by RAD7 and also by a cold-trap system for better accuracy.
The three radon sources and corresponding measurements are described as follows.

\begin{figure}[tb]
	\centering
	\begin{subfigure}[tph]{0.25\textwidth}
		\centering
		\includegraphics[height=1.5in]{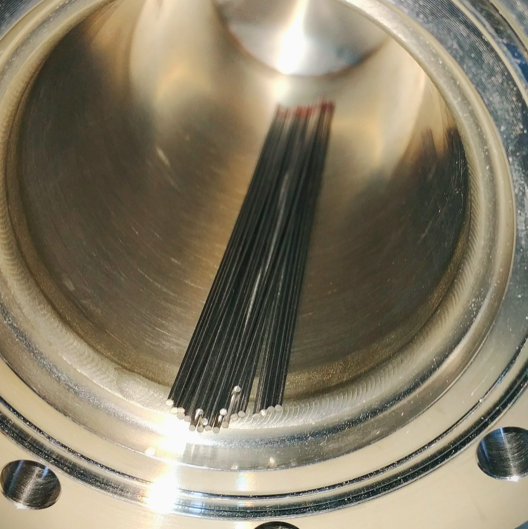}
		\caption{}
		\label{fig:electrodes}
	\end{subfigure}%
	~
	\begin{subfigure}[tph]{0.3\textwidth}
		\centering
		\includegraphics[height=1.5in]{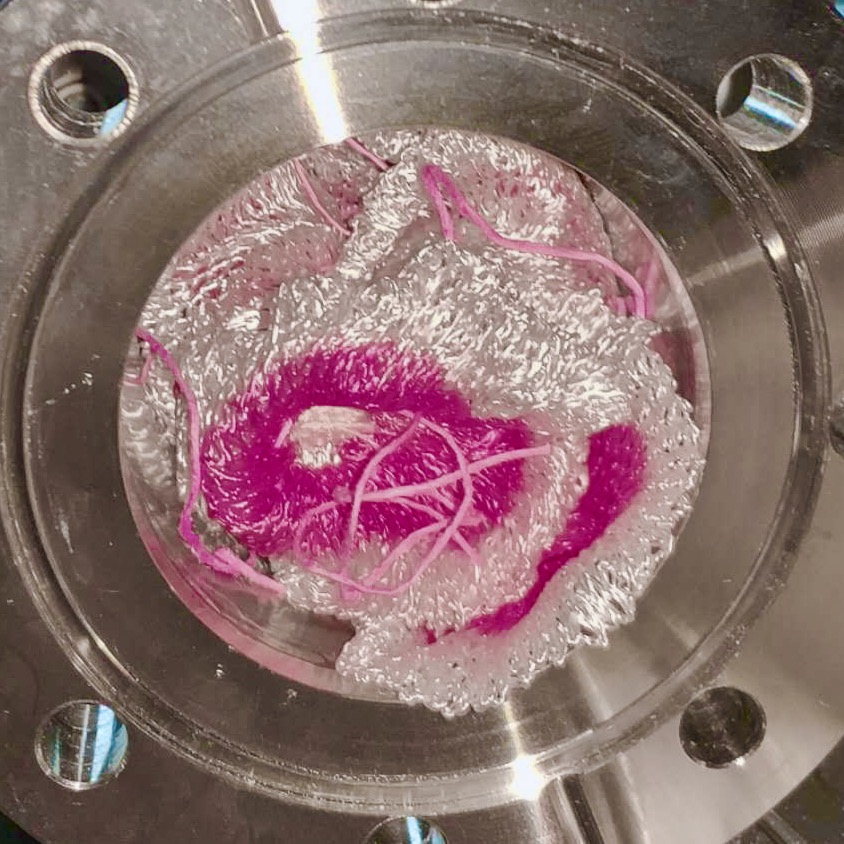}
		\caption{}
		\label{fig:mantles}
	\end{subfigure}
	~
	\begin{subfigure}[tph]{0.35\textwidth}
	\centering
	\includegraphics[height=1.5in]{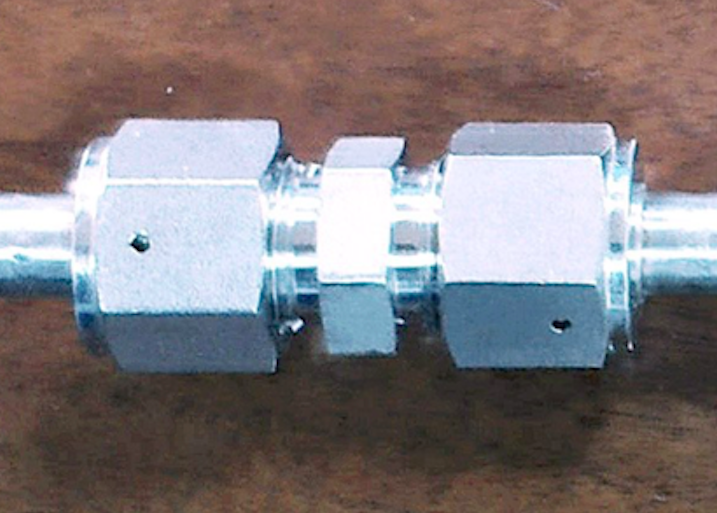}
	\caption{}
	\label{fig:sieve}
\end{subfigure}
\caption{
	Thoriated tungsten electrodes (a) and lantern mantles (b) are shown in the source chambers. Figure (c) shows the pipe male union where resin source were coated inside.
}
\end{figure}

Thoriated tungsten electrode with 1\% of thorium was used in the calibration campaign of 2017.
The diameter of the electrodes is about 1~mm and the length 150~mm.
A $^{228}$Th activity of 48.0$\pm$0.6~kBq/kg was measured in the HPGe detector with 109.7~g electrodes (20 pieces).
The $^{222}$Rn activity was less than $3.0$~kBq/kg at 90\% confidence level.
We measured emanation rates of 5~kg of electrodes for a continuous 25-hour duration with RAD7.
The emanated $^{220}$Rn and $^{222}$Rn activities were 1.22$\pm$0.03 and $0.03\pm0.01$~Bq/kg respectively.
The emanation rate was low, however the tungsten electrode source enabled us to study the behavior of $^{220}$Rn and its daughter isotopes in PandaX-II.

Lantern mantles treated with thorium nitrate (Th(NO$_3$)$_4$) were used as the radon sources in the 2018 campaign.
Commercial lantern mantles made from ramie fibers were used for their porous texture.
They were soaked in Th(NO$_3$)$_4$ solution and thoroughly dried for further usage.
The $^{228}$Th activity was measured to be $343\pm2$~Bq/piece, and $^{222}$Rn less than $0.58$~Bq/piece.
Ten pieces of lantern mantles were put into a 0.62-liter chamber for the emanation measurement.
A 0.5~$\mu$m particulate filter was added after the chamber to prevent any fiber from reaching RAD7.
With a continuous flushing of 24~hours, we measured the emanated $^{220}$Rn and $^{222}$Rn activities to be $1.27\pm0.01$ and $0.02\pm0.01$~Bq per piece, respectively.
In this calibration campaign, we collected sufficient number of low energy events to construct the ER background model for the dark matter analysis.

We conducted an additional $^{222}$Rn emanation measurement with a cold-trap collection system.
Instead of a direct measurement with RAD7, the source chamber was connected to a liquid nitrogen cold trap, where emanated $^{220}$Rn and $^{222}$Rn were frozen on the inner surface.
$^{220}$Rn would immediately decay while $^{222}$Rn accumulated on the surface.
The collection process in the cold-trap lasted for 6~hours and the frozen gas was transferred to a silicon photodiode detector setup for measurement.
The freeze-in efficiency of the cold trap and measurement efficiency of the silicon photodiode were determined by dedicated calibrations using a standard $^{222}$Rn source.
The silicon photodiode measured the temporal evolution of $^{218}$Po and $^{214}$Po activities during a 300-hour long measurement.
The fitted $^{222}$Rn activity was $3.6\pm 0.4$~mBq per piece.
We attributed the large discrepancy between RAD7 and cold trap measurement to possible overestimation from RAD7, in which high rate of $^{220}$Rn might introduce bias to the measurement of low rate of $^{222}$Rn.
It was also worth noting that we packed 60 pieces of lantern mantles for the cold trap measurement, which might negatively impact the emanation rates.

The source in the 2019 campaign emanated large amounts of $^{220}$Rn and $^{222}$Rn isotopes and were used to study event rates of $^{214}$Pb and its parents.
The source was fabricated by diffusing thorium nitrate into a thin layer of resin coated in a commercial pipe fitting, as seen in Figure~\ref{fig:sieve}.
Two 3~nm-rated filters were connected at the ends of the fitting.
Measured by the RAD7 detector, the $^{220}$Rn ($^{222}$Rn) activity was $300\pm3$ ($16\pm2$)~Bq.
The calibration was done right before the end of PandaX-II operation and therefore the high $^{222}$Rn activity was not a concern.

\section{Gaseous Source Delivery System}

\begin{figure}[tb]
	\centering
	\includegraphics[width=0.9\linewidth]{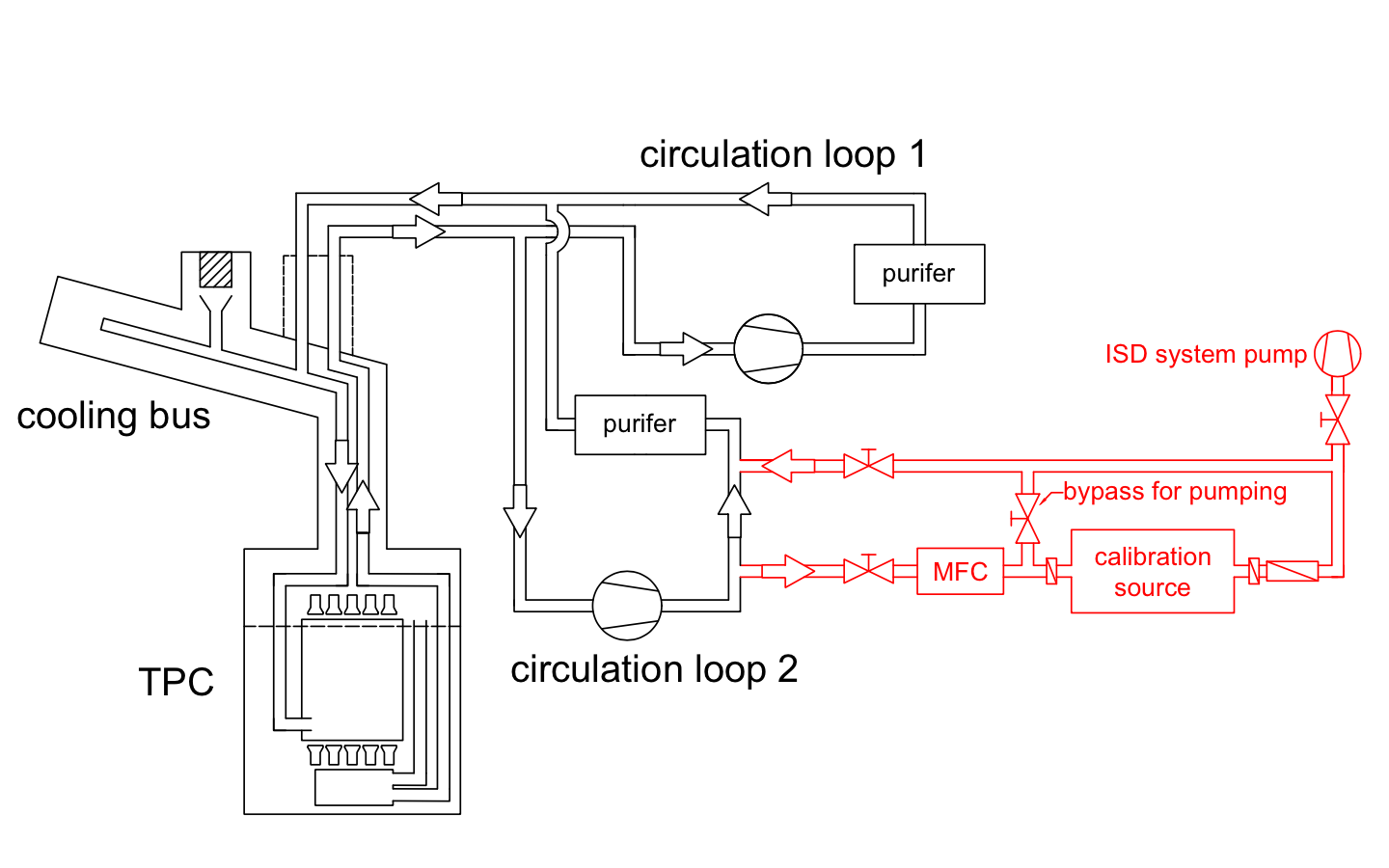}
	\caption{The gaseous source delivery system (in red) is shown with a simplified schematic of the PandaX-II detector and the xenon circulation and purification system~\cite{andiPhD}.
}
	\label{fig:sourcedeliverysysdiagram}
\end{figure}

We inserted radon sources into the PandaX-II circulation gas loop with an added set of delivery pipeline.
The central component of the system was a small chamber to contain the sources.
The auxiliary structure included a mass flow controller (MFC), a vacuum pump, a couple of bypass pipelines, valves, and filters, as shown in Figure~\ref{fig:sourcedeliverysysdiagram}.
The source chamber was a stainless steel cylinder with standard Conflat flanges on the two ends.
The nominal diameter was 63~mm and length 200~mm.
The Conflat flanges were coupled to the pipeline with 1/4-inch VCR fittings.
All chambers, flanges, and pipes were electrochemically polished to reduce possible $^{222}$Rn emanation.
A 0.4~$\mu$m-rated gasket-type filter and a 3~nm-rated filter were mounted downstream of the source chamber to prevent radioactive particulates from entering the gas loop.
A 0.4~$\mu$m-rated filter upstream was also installed to avoid any negative impact when the gas flow might have been paused.

We mounted the setup in parallel with the existing circulation loops (Figure~\ref{fig:sourcedeliverysysdiagram}).
The PandaX-II detector was connected to two parallel circulation loops, in which xenon gas was continuously purified through high temperature zirconium getters at a typical flow rate of 40 to 60~SLPM.
Liquid xenon in the detector was extracted and warmed up to gas phase before purification.
Purified gas was then liquefied by a heat exchanger and a dedicated cooling module before fed into the TPC again.
For calibration campaigns, the source module was placed upstream of the getter on loop 2 to minimize the impact of possible impurity gas emanated from the $^{232}$Th source.
The amount of radon reaching the TPC depended on the flow rate through the source as well as the overall flow rate of circulation xenon gas.
During the three calibration campaigns, the gas flow rate through the source chamber was about 20, 4, and 1~SLPM respectively.

\section{Performance in the TPC} \label{sec:performance}

We performed three radon calibration campaigns from 2017 to 2019.
Detector performance, including trigger rate,  electron lifetime, and  DAQ system deadtime, was carefully monitored during the source injections.

Initial trial injections in 2017 caused drift electron lifetime to decrease to about 100~$\mu$s in several runs, possibly due to impurity in the injection pipelines flushed to the detector.
The detector quickly recovered and then we resumed calibration runs without further interruption.
The flow rate through the source chambers in the other two campaigns were much smaller than that of 2017 and therefore negligible impact was introduced to the TPC.

In 2018, we introduced a large amount of $^{220}$Rn and the DAQ system get saturated with trigger setting the same as in dark matter search runs.
We implemented a new set of calibration trigger to record low energy ER events efficiently while suppressing high energy alpha events.
We maintained trigger rates were around 32~Hz during our calibration runs.
To measure the event rates of $\alpha$s, we took data for 43 minutes with a random trigger of about 120~Hz and each triggered event recorded 1~ms-long waveforms of all PMTs.

We characterized the event rate changes of $^{220}$Rn and its daughters in the detector over time, and examined possible permanent $^{232}$Th contaminations and $^{222}$Rn contaminations during and after the injection period.
The results are presented in subsection~\ref{sec:evolution} and detailed event selection procedures described in subsection~\ref{sec:alpha} to \ref{sec:coincidence}.

For the final radon calibration in May 2019, the emphasis was on $^{222}$Rn.
About $1$~Bq of $^{222}$Rn was accumulated in detector in two days before we stopped injection.
We were able to collect a large sample of $^{214}$Pb and $^{214}$Bi-$^{214}$Po coincidence events for another 5~days after injection.
The ratio of event rates of $^{218}$Po and $^{214}$Bi-$^{214}$Po is reported in subsection~\ref{sec:pb214}.

\subsection{Event Rate Evolution and Calibration-Induced Background} \label{sec:evolution}

\begin{table}[tb]
	\centering
\begin{tabular}{p{0.25\textwidth}ccc}
\toprule
Campaigns & 2017 & 2018 & 2019 \\
\hline
Thoriated Source & Tungsten electrodes & Lantern mantles & Coated resin \\
Duration [days] & 35 & 20 & 2.1\\
Live Time [days] & 18.9 & 11.9 & 1.3\\
Injected {$^{220}$Rn} [Bq] & $2.521\pm0.001$ & $31.7\pm 0.3$ & $45.5\pm0.2$ \\
Injected {$^{222}$Rn} [Bq]& $(1.60\pm0.02)\times 10^{-2}$ &  $0.28\pm0.04$ & $6.1\pm0.1$ \\
{$^{220}$Rn} from thorium particulate [$\mu$Bq/kg] & 0.20$\pm$0.03  & -0.01$\pm$0.06 \\
\bottomrule
\end{tabular}
	\caption{
The activity of injected $^{220}$Rn and $^{222}$Rn were calculated.
$^{220}$Rn from thorium particulate refers to the possible contaminations induced by thorium compound particulate reached the TPC active volume. The value was measured by the difference between $^{220}$Rn activity before and after the radon calibration campaigns.
The activities after injections were measured after the injected $^{220}$Rn had decayed to insignificant levels.
}
	\label{tab:pollution-table}
\end{table}
\begin{figure}[tb]
	\centering
	\includegraphics[width=0.75\linewidth]{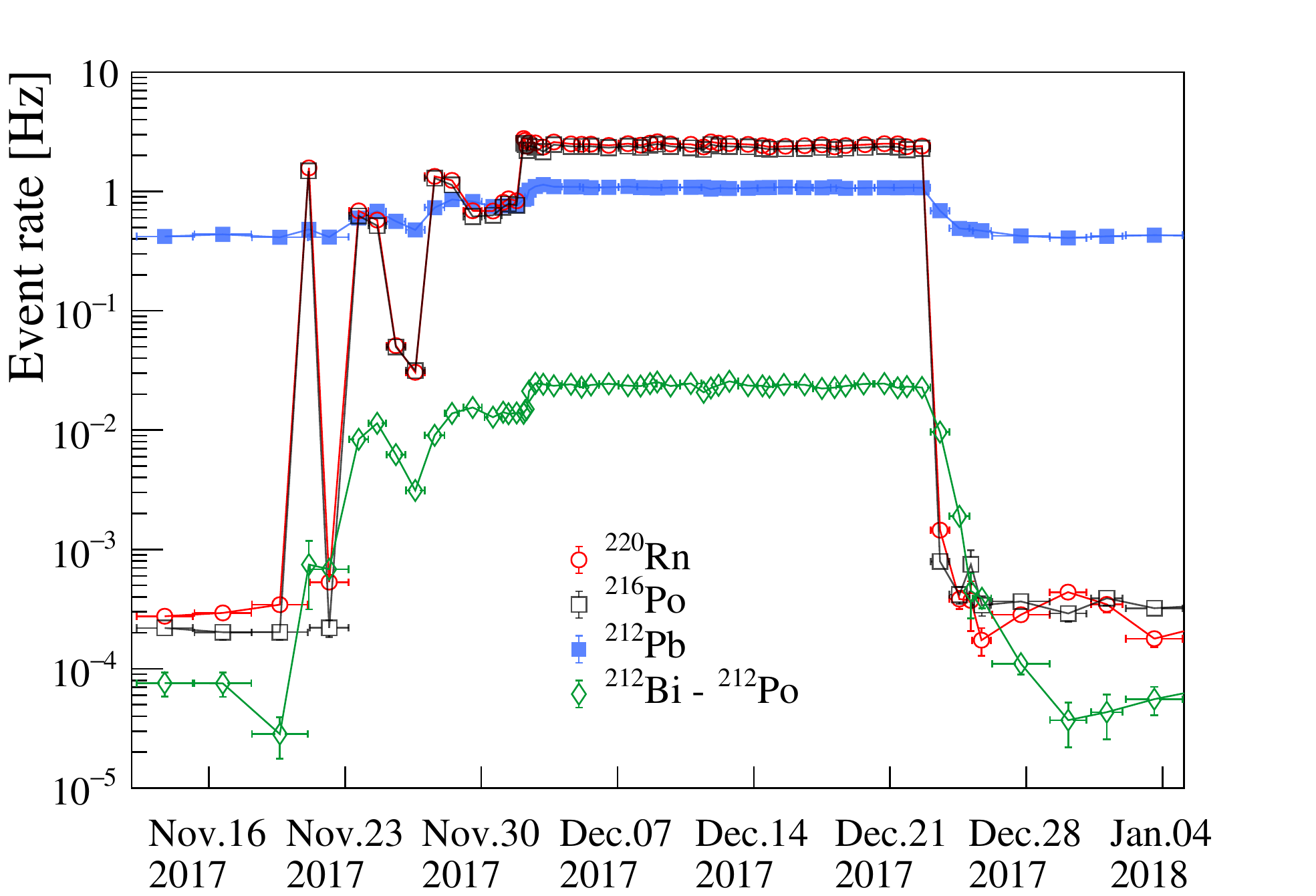}
   \caption{
	The event rate evolution of $^{220}$Rn and daughters during the 2017 calibration.
	The vertical error bar refers to the statistical uncertainty and the horizontal bar represents the run duration.
	Event rates of $^{220}$Rn and $^{216}$Po are from $\alpha$s, $^{212}$Pb from events in the energy range of [0, 600]~keV$_{ee}$, and $^{212}$Bi-$^{212}$Po from the coincidence algorithm in~\ref{sec:coincidence}.
	Spikes in the event rates in November are due to test injections.
}
	\label{fig:evo17}
\end{figure}
Event rates during calibration were used to characterize the accumulation, equilibration, and decay of the gaseous source in the TPC.
We measured rates from high energy $\alpha$ events from $^{220}$Rn and $^{216}$Po, low energy $\beta$ events from $^{212}$Pb, and $\beta$ and $\alpha$ coincidence events from $^{212}$Bi and $^{212}$Po respectively.
In Figure~\ref{fig:evo17}, we showed the temporal evolution of the rates during the 2017 calibration campaign.
For 2017, the duration of source injection campaign lasted 35 days, during which we were able to accumulate 16.9 days of live data with stable $\alpha$ event rates.
We performed a few trial injections in late November 2017 and we could see $\alpha$ event rate spikes in the figure.
$^{220}$Rn rate increased by $2.521\pm0.001$~Hz during stable source injection comparing with background data before injection.
The event rate changes were more drastic in 2018 and 2019, as shown in Table~\ref{tab:pollution-table}.

A smaller amount of $^{222}$Rn was also introduced into the TPC by the injection.
The ratio between event rates of injected $^{222}$Rn and $^{220}$Rn for 2017 (2018) was $0.63\pm0.01~\%$ ($0.90\pm0.12~\%$).
The small ratios here meant that introduced $^{222}$Rn had only a manageable contribution to dark matter data taking.

The other major concern for the radon sources was permanent contamination by particulate thorium compound infiltration.
Due to the long half-life, $^{232}$Th and daughter isotopes would slowly but steadily release $\alpha$, $\beta$, and $\gamma$ background from inside of the TPC.
We quantified the effect by comparing $^{220}$Rn event rate differences before and after the calibration campaigns.
The measured activities, normalized by fiducial volume, are given in Table~\ref{tab:pollution-table}.
The result from the 2018 campaign was consistent with zero and showed that no significant amount of thorium particulate was introduced to the detector.

We stopped PandaX-II operation after the radon calibration campaign in 2019, and no measurement was done about particulate contaminations.

\subsection{High Energy $\alpha$ Spectrum}
\label{sec:alpha}

\begin{figure}[tb]
	\centering
\includegraphics[width=0.75\linewidth]{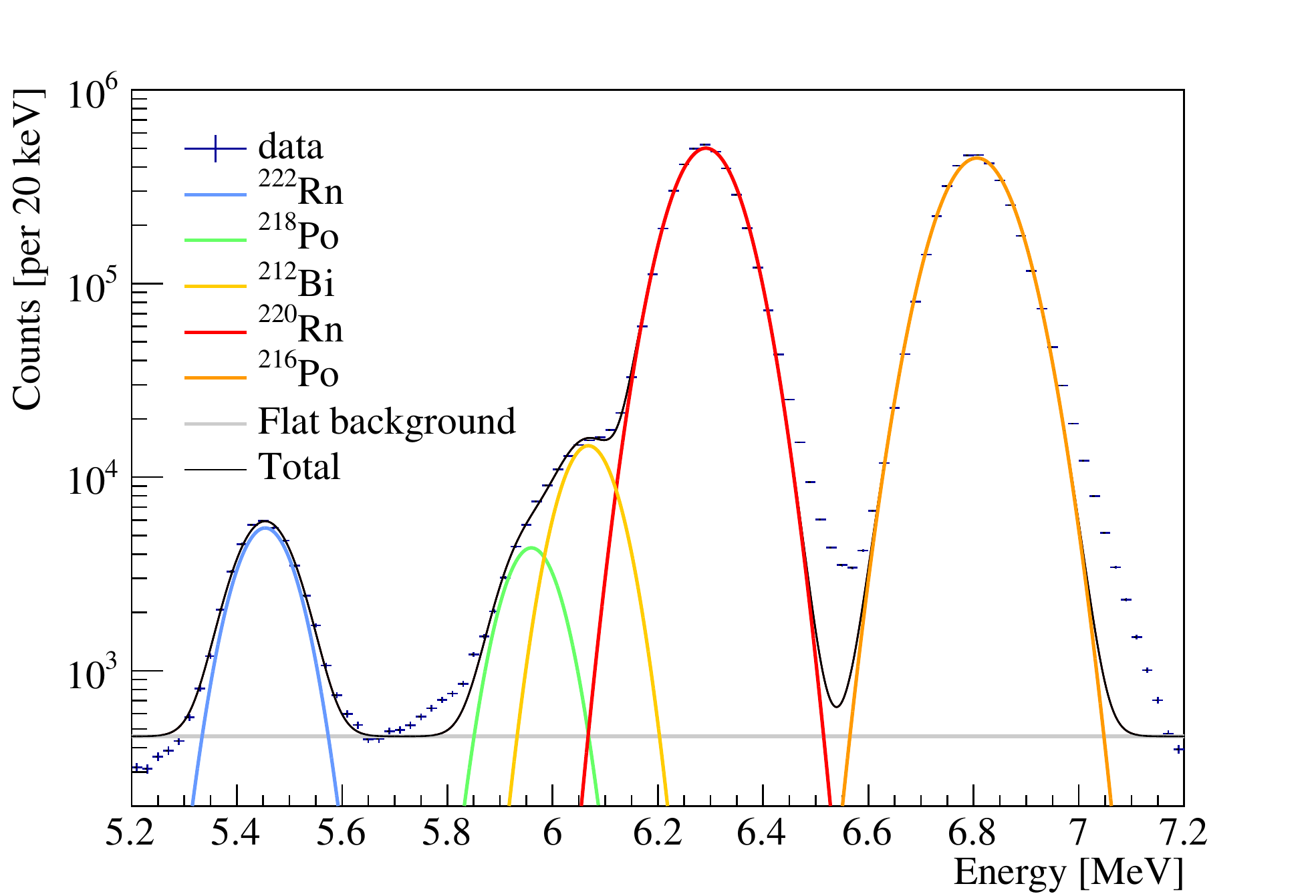}
	\caption{High energy region of the spectrum for calibration data in 2017. $^{222}$Rn (5490 keV), $^{218}$Po (6002 keV), $^{212}$Bi (6090 keV), $^{220}$Rn (6288 keV), and $^{216}$Po (6778 keV) peaks are fitted with Gaussian functions and drawn in blue, green, yellow, red, and orange, respectively.
A flat background is added in the fitting, as shown in gray.
The left shoulder of $^{218}$Po and right-hand sides of the $^{220}$Rn, $^{216}$Po peaks deviate from fitting due to response distortion for $S1$ signals near the bottom of the detector.
}
	\label{fig:ec1}
\end{figure}

High energy $\alpha$ events from $^{220}$Rn, $^{222}$Rn, and their progenies were selected with spatial and energy cuts.
Events in the top and bottom 17~mm-thick layer were rejected with drift time cuts.
$\alpha$ events close to the edge of the field cage, which we called surface events, were rejected by requiring $S2>25000$~PE.
We also removed events with incorrectly reconstructed z-positions by comparing expected and measured relative light collection between the top and bottom PMT arrays.
A $\beta$ event and other low energy event fallen in the same acquisition window as $\alpha$ might bias the reconstructed energy.
We identified those coincident events by the $S1$ amplitude and rejected any events with $S1$ outside the $[30000, 70000]$~photoelectron (PE) range.
After 3D mapping correction (see below), we further selected $\alpha$ events with more strict $S1\in [40000, 60000]$~PE and $S2>30000$~PE cuts.

Energy of $\alpha$ events was reconstructed with scintillation $S1$ signals after three-dimensional position corrections.
The position of each event was reconstructed from $S1$ and $S2$ signals.
A full-volume mapping of $S1$ response was generated from 100 million $^{220}$Rn-$^{216}$Po coincidence events.
$S1$ signals were subsequently corrected taking into account the non-uniformity in detector response.
The corrected $S1$ signals were fitted linearly with signature $\alpha$ peak energies and the resulting light yield was $L_y = 8.19\pm0.10$~PE/keV.
We fitted the peaks with Gaussian functions and a flat background in Figure~\ref{fig:ec1}.
Fitted counts were used to report event rates of radon isotopes during the calibration runs.
Deviation from Gaussian were seen on the left shoulder of $^{218}$Po and  right-handed sides of the $^{220}$Rn and $^{216}$Po peaks.
We confirmed the deviations were caused by events nearby the bottom of the TPC, where detector response to charges of large $S1$ signals was distorted.

\subsection{Low Energy $\beta$ Spectrum}
\label{sec:beta}
\begin{figure}[tb]

	\centering
		\centering
		\includegraphics[width=0.75\textwidth]{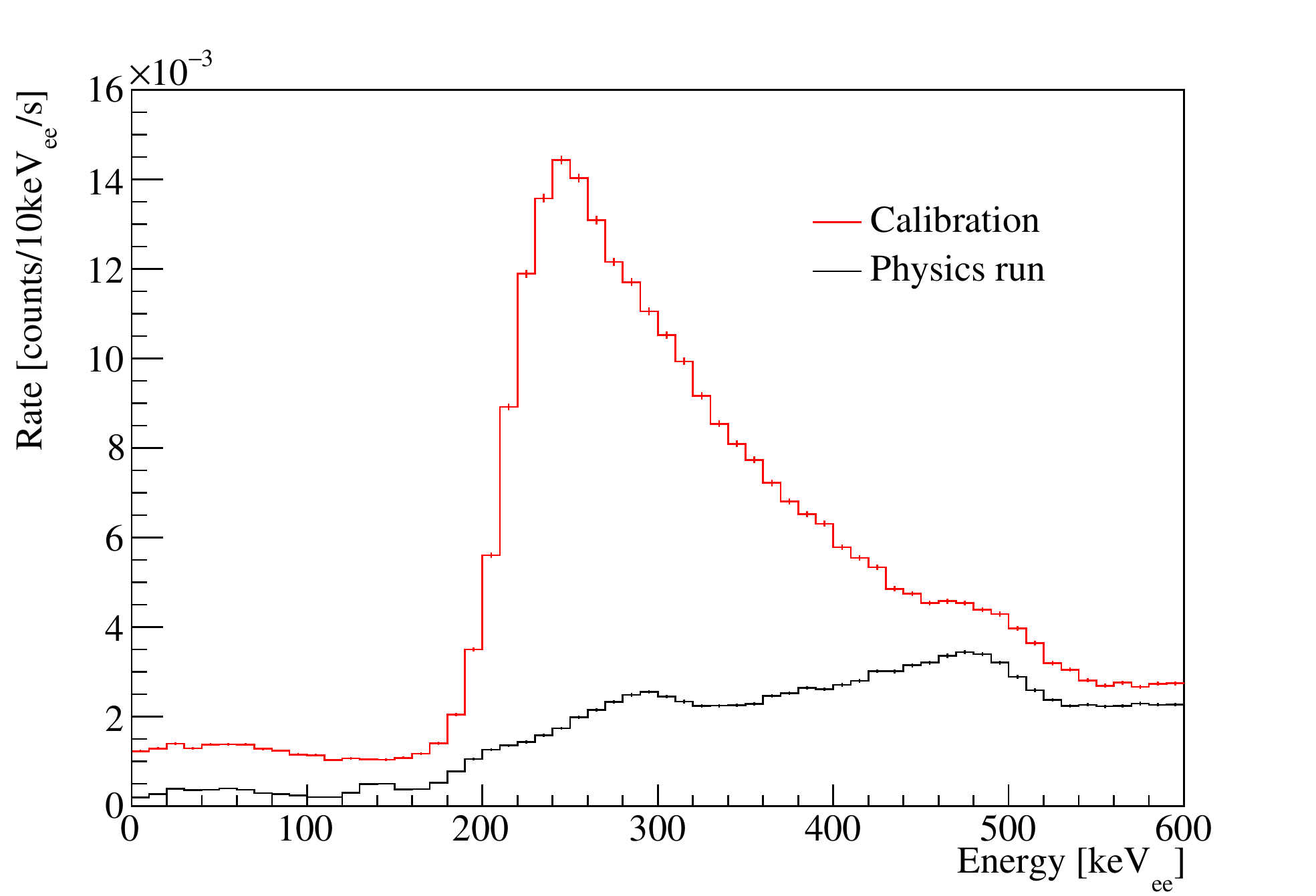}
	\caption{
Energy spectra in the range of 0 to 600~keV$_{ee}$ during calibration (red) and physics runs (black) in 2017.
For the physics (calibration) runs, a total of 427 (406) hours' data were used.
}
\label{fig:pb212data}
\end{figure}
\begin{figure}[tb]
	\centering
		\includegraphics[trim=0cm 0cm 2.5cm 1.5cm, clip=true,height=1.4in]{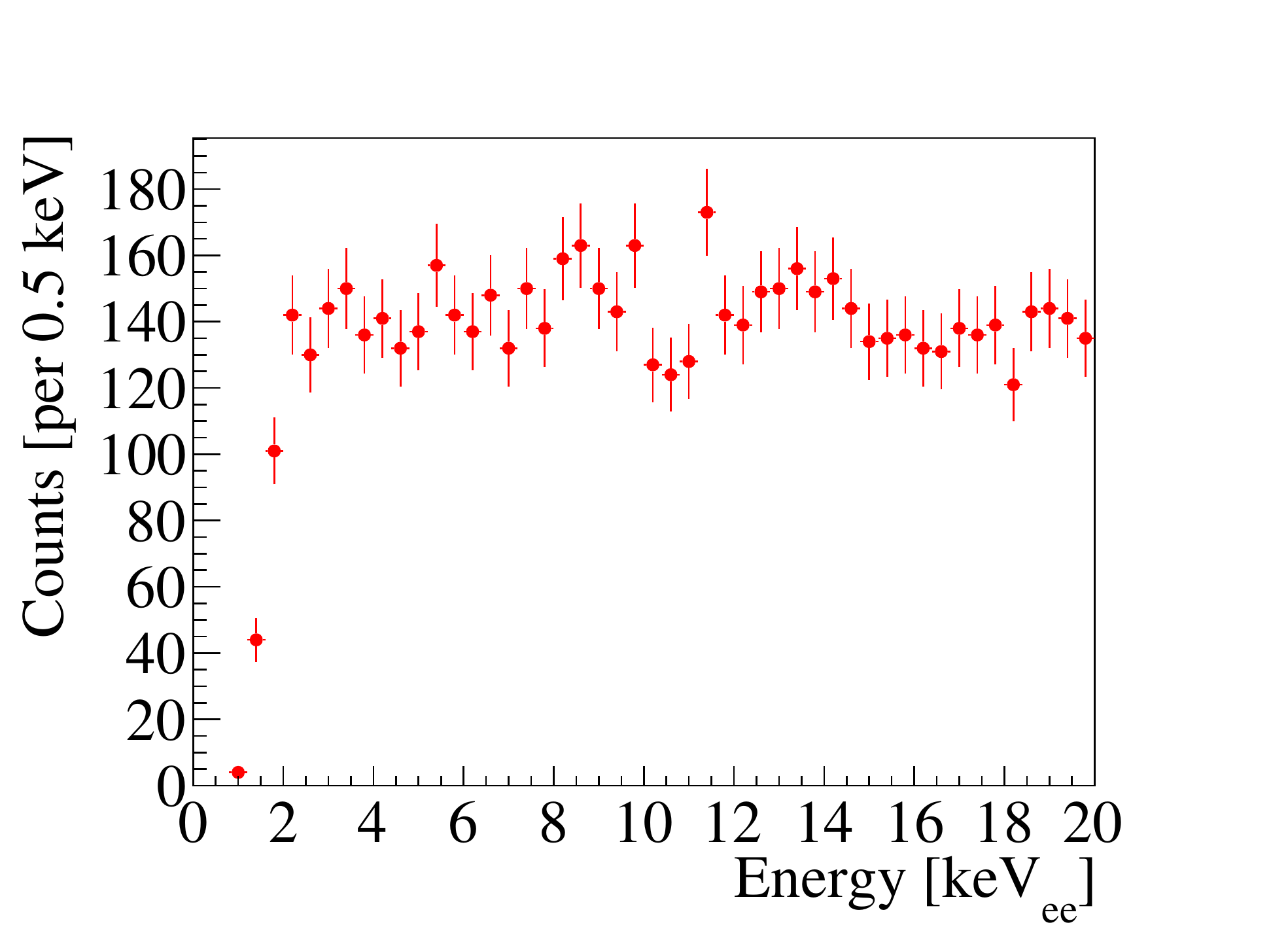}
		\includegraphics[trim=0cm 0cm 2.5cm 1.5cm, clip=true,height=1.4in]{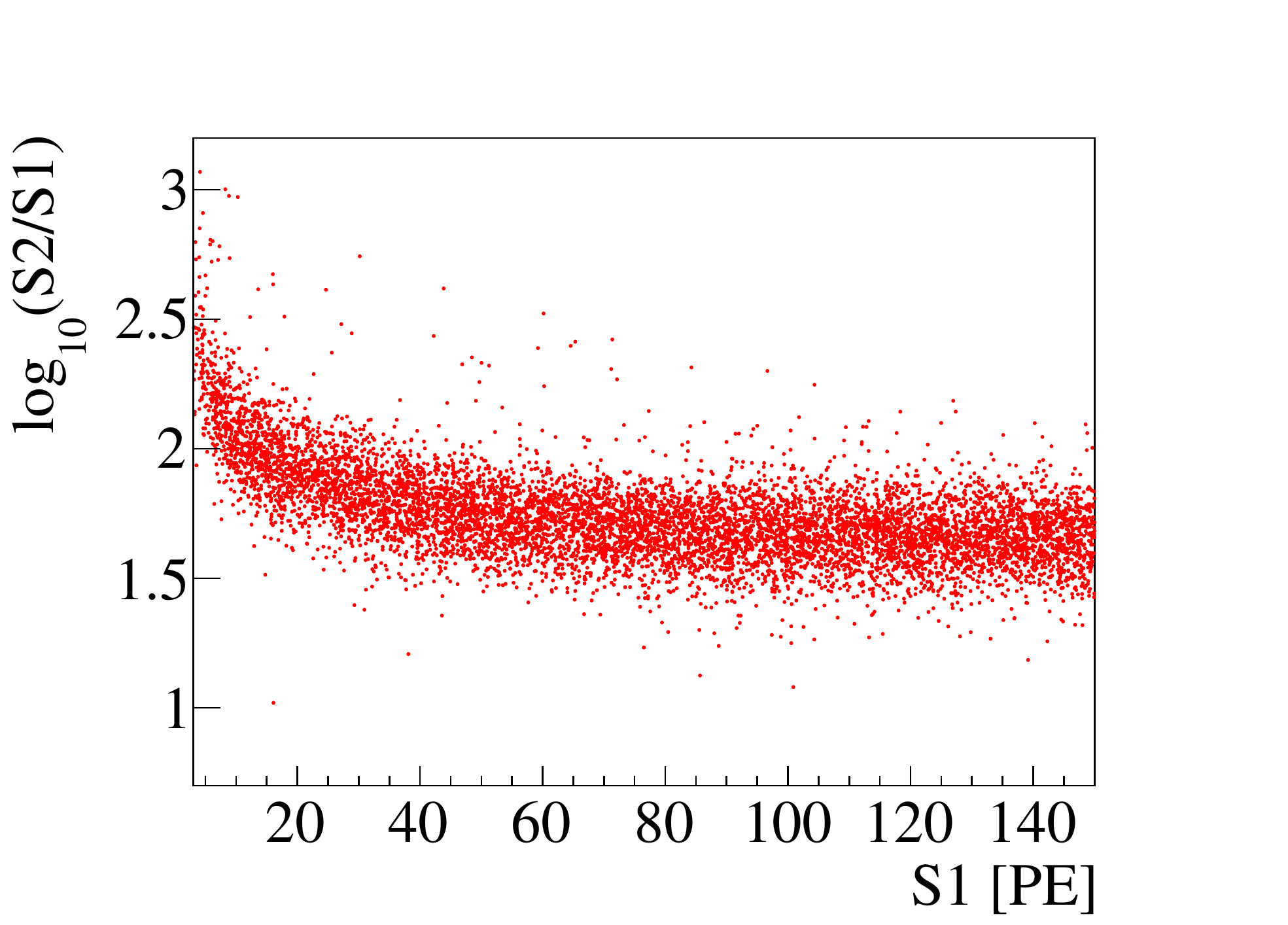}
		\includegraphics[trim=0cm 0cm 2.5cm 1.5cm, clip=true,height=1.4in]{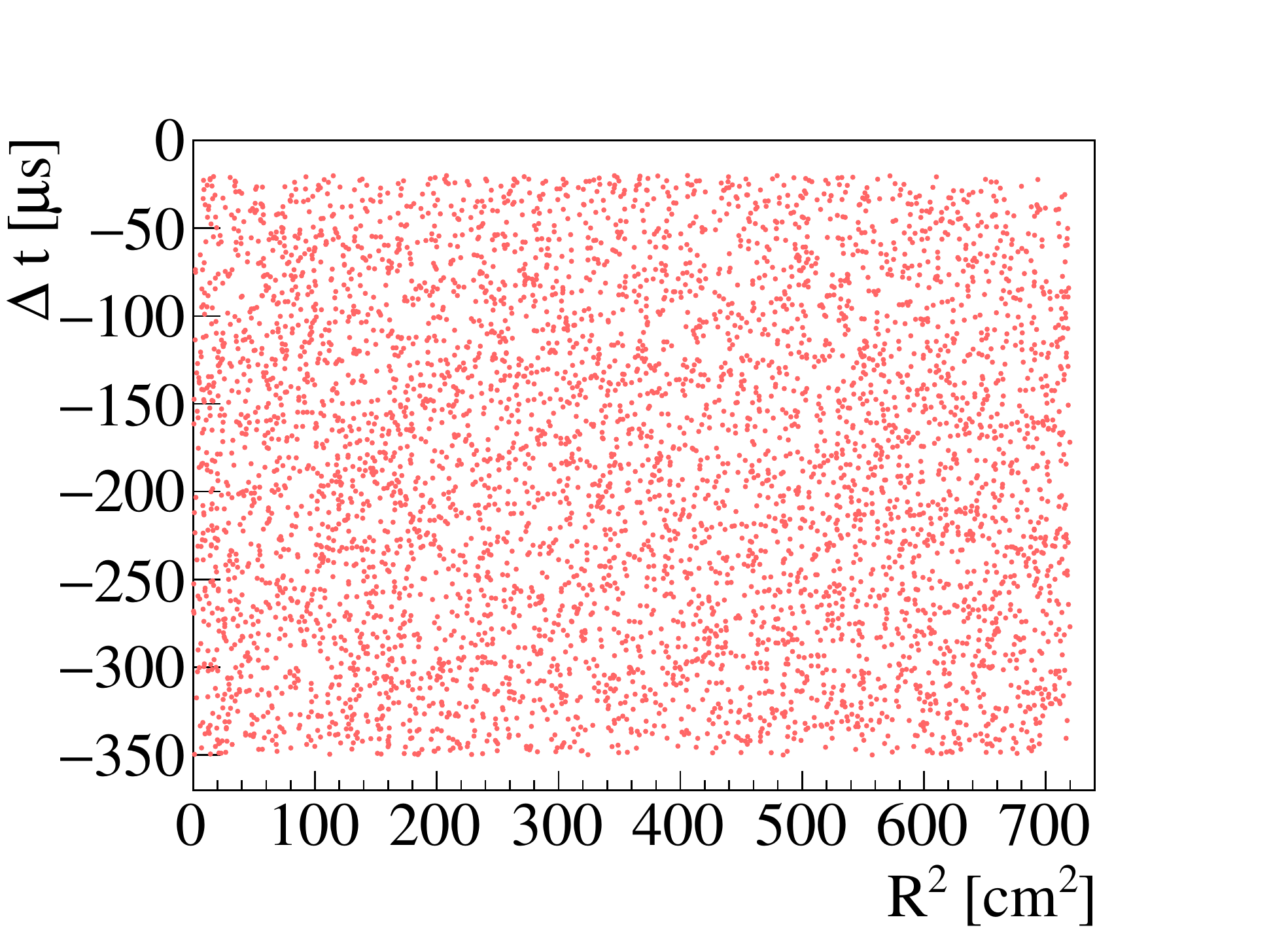}
	\caption{
		Reconstructed energy spectrum (left) and ER band (middle) and spatial distribution (right) obtained from the calibrations run in 2018.
		Events were selected with identical fiducial volume cuts and pulse quality cuts as in Ref.~\cite{final-analysis,Zhou:2020bvf}.
	}
\end{figure}
We monitored the event rate change of the low energy region in the range of $[0,600]$~keV$_{ee}$.
During the calibration runs, this part of the spectrum was dominated by $^{212}$Pb events, which were typical multi-scattering events of $\beta$ and $\gamma$ particles from the $^{212}$Pb decay.
The electron equivalent recoil energy is reconstructed as the linear combination of the $S1$ signal and the sum of multiple $S2$ signals.
Energy reconstruction parameters such as photon detection efficiency (PDE), electron extraction efficiency (EEE), single electron gain (SEG), and low-energy efficiencies were inherited from~\cite{final-analysis}.
Among them, PDE and EEE were determined from multiple mono-energetic $\gamma$ lines in the energy range.
Figure~\ref{fig:pb212data} shows the low energy $\beta$ spectra before and during the calibration runs of 2017 in the volume that $\Delta t \in [20,350]\, \mu s$ and $x^2+y^2<720$~cm$^2$.
The $\beta$ and the subsequent 238~keV $\gamma$ were captured by the TPC almost simultaneously and the spectrum clearly shows the signature.

Low energy electron recoil data from 2018 calibration were used for the ER model in the PandaX-II full-exposure analysis~\cite{final-analysis}.
Besides fiducial volume cuts and quality cuts, we introduced an additional quality cut to reject events with excessive noise in pulses, which was also adopted in the dark matter analysis.
In the range of $S1\in[3,150]$~PE, 8973 events were obtained after all cuts during 11.9 days of live time.
The average rate of the observed low energy events was $754\pm8$ events per day, which was only about 2.7\% of the expected if we assumed secular equilibrium between $^{212}$Pb and $^{220}$Rn and took into account an energy acceptance of 0.8\% in $[0,20]$~keV for $^{212}$Pb events.
The small $^{212}$Pb rate could be attributed partly to the mobility of charged ions drifting towards the cathode and out of the fiducial volume~\cite{EXOmobility, RnKrXENON100}.

\subsection{Delayed Coincidence Events} \label{sec:coincidence}

\begin{figure}[tb]
\centering
\includegraphics[width=0.75\linewidth]{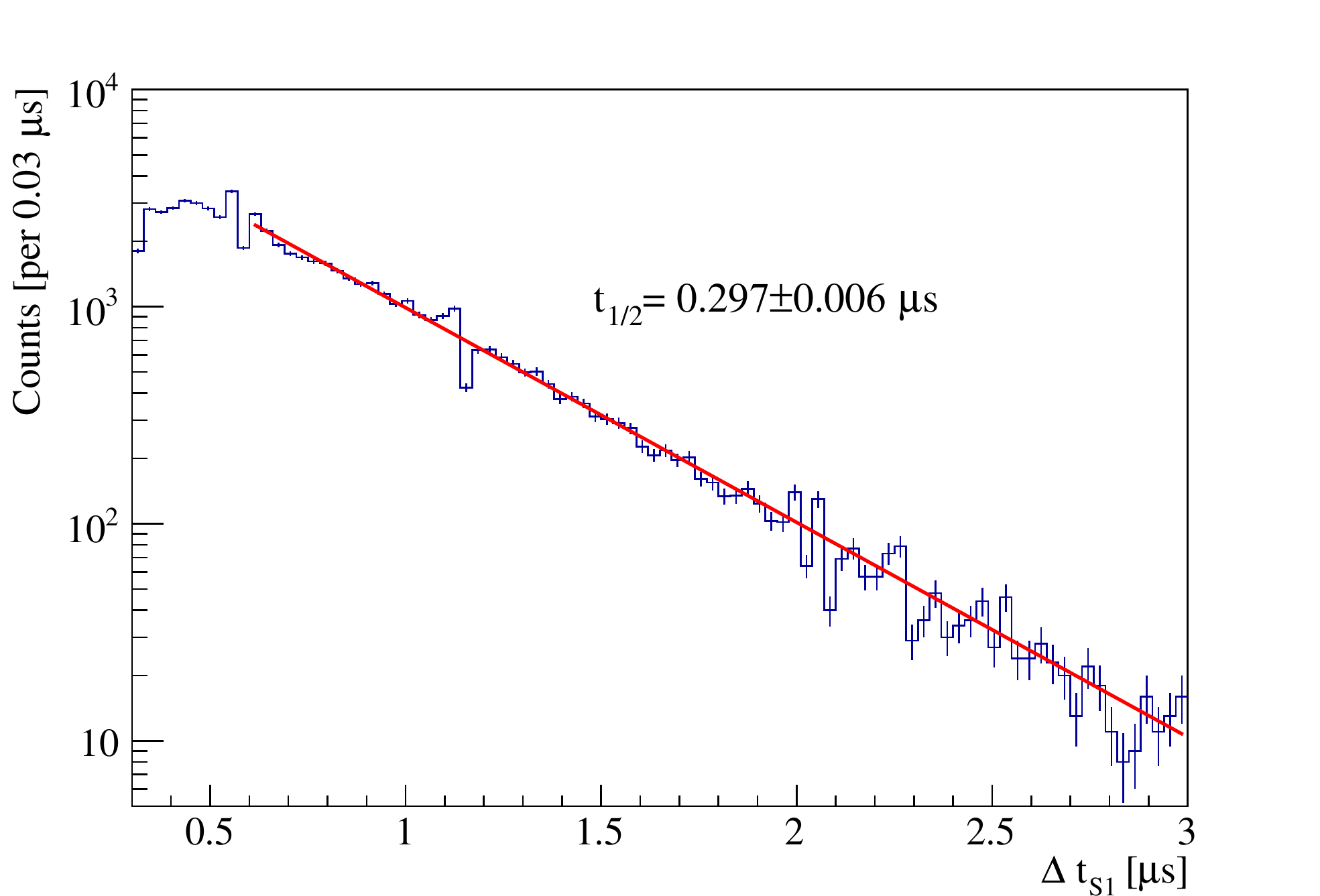}
\caption{
	The time difference between decay events of $^{212}$Bi and $^{212}$Po.
	Deviation from exponential with smaller $\Delta t_{S1}$ is due to the inefficiency of separating two $S1$ signals.
}
\label{fig:po212lifetime}
\end{figure}

$^{212}$Bi-$^{212}$Po events in Figure~\ref{fig:evo17} refer to the coincidence events of a $\beta$ from $^{212}$Bi (end point 2254~keV) and an $\alpha$ from $^{212}$Po (8784~keV), with a typical delay time scale of $0.3~\mu$s.
The coincidence events had a typical double-$S1$ pattern and high energy depositions from $\alpha$s and thus could be easily identified.
$^{214}$Bi-$^{214}$Po event selection used in the next section followed similar algorithm.

Selection cuts were implemented specifically for coincidence events and corresponding efficiencies were calculated.
The $S1$ signal of $^{212}$Bi $\beta$ was firstly selected to have energy in the range of 50 to 3000 keV$_{ee}$, which introduced a 98.1\% efficiency \cite{Shaoli}.
The events were selected in the bulk volume of the detector.
Electrode events were rejected by requiring $\Delta t \in [10,350]$~$\mu s$.
Fiducial cuts as well as $\alpha$-specific surface event cuts were applied to selected events in the bulk of the detector with the same cut values as mentioned previously.
$S2$ signals of $\beta$ and $\alpha$ events are often overlapped since their width is on the scale of microseconds.
Therefore, the position of the events was reconstructed from the timing difference between $S2$ and $S1_{\beta}$.
Random coincidence events were rejected by the top-bottom charge ratio cut, which checked the correlation between the ratio and event z-coordinate.
The timing difference between $S1$ signals of $\beta$ and $\alpha$ $\Delta t_{S1}$ of events is plotted in Figure~\ref{fig:po212lifetime}.
An exponential decay curve could fit data with  $\Delta t_{S1}$ larger than 0.6~$\mu$s.
The fitted half-life $t_{1/2}=0.297\pm0.006\,\mu$s agreed with the expected value.
And the uncertainty was dominated by systematic uncertainty from fitting range choices and fiducial volume.
The curve deviated from exponential at smaller $\Delta t_{S1}$ when $S1_{\alpha}$ and $S1_\beta$ could not be effectively separated.
The efficiency of this effect was estimated to be 75\%.

\subsection{$^{214}$Pb measurement from $^{222}$Rn Injection}
\label{sec:pb214}
\begin{figure}[tb]
	\centering
	\includegraphics[width=0.7\linewidth]{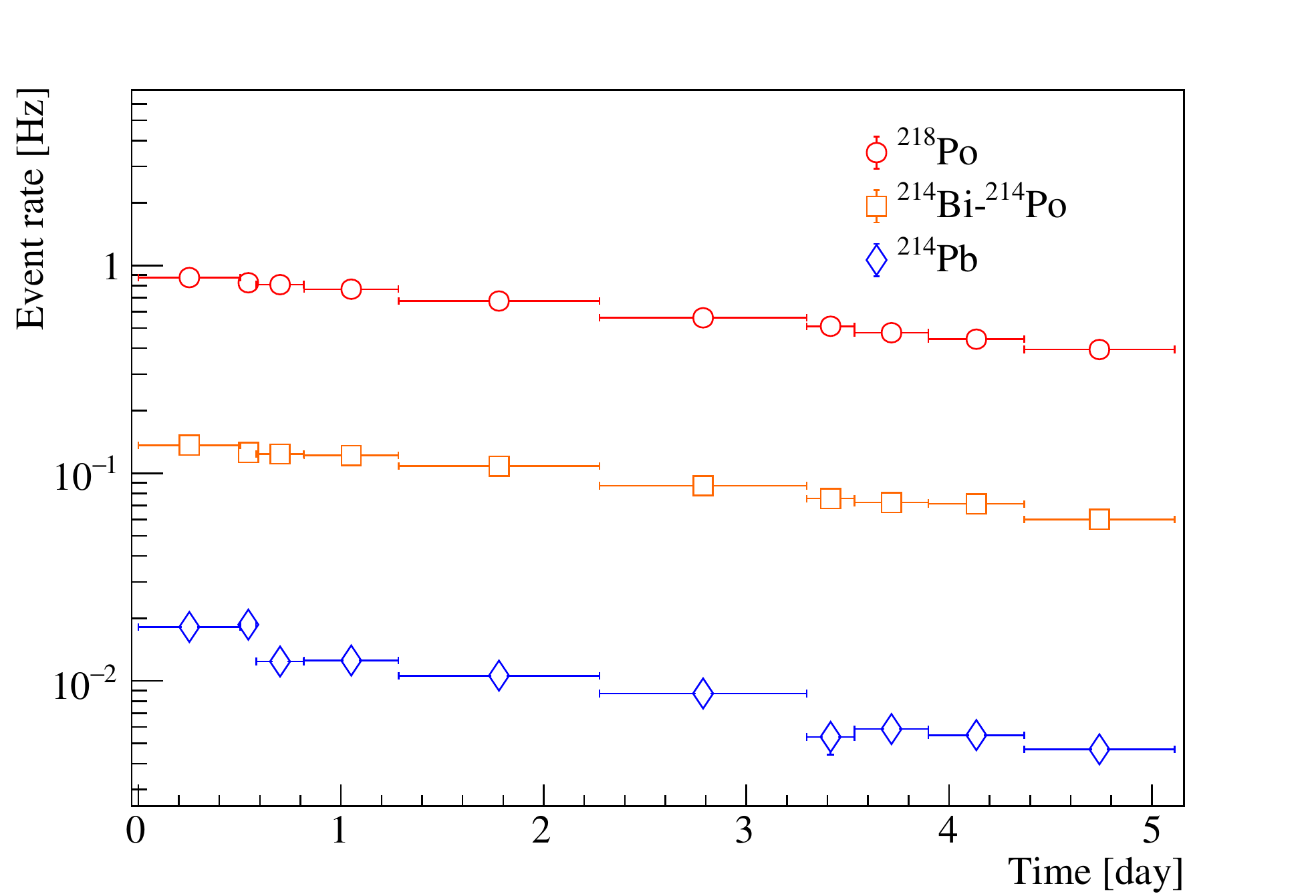}
   \caption{
	Event rate evolution of $^{214}$Pb and parent isotopes in 2019.
The starting time in the horizontal axis is about 35 hours after the source injection stopped.
	$^{218}$Po rates are calculated from $\alpha$s and $^{214}$Bi-$^{214}$Po rates from the coincidence algorithm.
	$^{214}$Pb rates are calculated with events in the range of [10,200]~keV$_{ee}$.
	Background is subtracted for all rate calculations.
}
	\label{fig:pb214ratio}
\end{figure}
A common difficulty in the spectrum fitting in dark matter searching data was the contribution from $^{214}$Pb in the low energy range.
The low energy ER background contribution of $^{222}$Rn and its progenies was dominated by  $^{214}$Pb.
While the radioactivities of ancestors on the decay chain, such as $^{222}$Rn, can be estimated from high energy events, it cannot be used to estimate $^{214}$Pb activity reliably.
A decrease of activities when compare parent to daughter isotopes is often observed and we call this phenomena chain depletion.
A quantitative evaluation of chain depletion is desirable to better constrain the $^{214}$Pb contribution in the low energy spectrum.
In PandaX-II, we measured the chain depletion factor directly by comparing the event rates of $^{214}$Pb and $^{218}$Po as well as  $^{214}$Bi-$^{214}$Po and $^{214}$Pb with the help of a $^{222}$Rn source.

The thoriated resin calibration source used in the 2019 campaign injected a large amount of $^{222}$Rn in the TPC and provided us a rare opportunity to study the chain depletion.
In the first week after calibration was stopped, $^{220}$Rn and its daughters quickly decayed to insignificant levels and $^{214}$Pb events dominated the energy region up to 1000~keV$_{ee}$.
The activity of $^{214}$Pb in the TPC was determined based on event rates in the energy range [10, 200]~keV$_{ee}$.
After subtracting backgrounds, the event rate was then divided by 4.3\%, the normalization factor of events in [10, 200]~keV$_{ee}$ with respect to the full $^{214}$Pb spectrum based on a GEANT4 simulation~\cite{geant4-2016, geant4-2006}.
Delayed coincidence events of $^{214}$Bi-$^{214}$Po were selected by the cuts outlined in the previous subsection.
The event rates of $^{218}$Po, $^{214}$Pb, and $^{214}$Bi-$^{214}$Po are as shown in Figure~\ref{fig:pb214ratio} as a function of time.
The chain depletion ratio between $^{214}$Pb and $^{218}$Po was determined to be $36.6\%\pm1.0\%$, and ratio between $^{214}$Bi-$^{214}$Po coincidence events and $^{214}$Pb was $43.1\%\pm1.4\%$.
These ratios enabled us to make a precise data-driven estimate of the low-energy ER background contributed by $^{214}$Pb.
This new procedure was applied in the PandaX-II final dark matter analysis in Ref.~\cite{final-analysis}.
The rate of background event from $^{214}$Pb, determined from depletion ratio and measured $\alpha$ rate of $^{218}$Po or $^{214}$Bi-$^{214}$Po coincidence rate, was used as input initial value and constraint in the spectrum fitting in search of dark matter.

\section{Conclusion}

Internal calibration sources are of increasing importance with the larger liquid xenon detectors being commissioned and planned.
We investigated three $^{220}$Rn emanation sources based on natural thorium compounds.
All the sources have been tested in the PandaX-II detector.
By flushing thoriated tungsten electrodes and lantern mantles, we were able to introduce a significant amount of $^{220}$Rn without serious $^{222}$Rn contaminations.
We also observed no thorium particulate infiltration in the TPC with lantern mantles, which is especially important to avoid introducing any permanent contaminations.
The two sources are made of commercial products and can be easily prepared.
A large sample of low energy ER calibration events, mainly from $^{212}$Pb, were collected and used to construct a data-driven ER response model for the dark matter searches of the PandaX-II.
The thoriated resin source were used mainly as a $^{222}$Rn source to measure the activity ratios between $^{214}$Pb and its parent isotopes.
This allows a robust estimate of $^{222}$Rn-induced background in the low energy dark matter search region.

We are investigating the possibility to use similar natural thorium compound sources for the PandaX-4T detector, which is currently under construction.
Even by conservatively assuming that the ratio between $^{222}$Rn and $^{220}$Rn for PandaX-4T calibration is the same as what we observed in 2018, it would take around 25 days of waiting time before we could resume dark matter data taking.
However, we envision that calibrations for other purposes, such as neutron calibrations, can be performed during the $^{222}$Rn cooling-off period.
With this strategy, we can minimize the impact to overall physics run time.

\section*{Acknowledgement}
We thank Shoukang Qiu and Quan Tang from University of South China for helping fabricate the thoriated resin source.
This project is supported by grants from the Ministry of Science and Technology of China
(No. 2016YFA0400301 and 2016YFA0400302), a Double Top-class grant from Shanghai
Jiao Tong University, grants from National Science Foundation of
China (Nos. 11435008, 11505112, 11525522, 11775142 and 11755001), grants from the Office of Science and Technology, Shanghai Municipal Government (Nos. 11DZ2260700, 16DZ2260200, and 18JC1410200), and the support from the Key Laboratory for Particle Physics, Astrophysics and Cosmology, Ministry of Education.
This work is supported also by the Chinese Academy of Sciences Center for Excellence in Particle Physics (CCEPP) and Hongwen Foundation in Hong Kong.
Finally, we thank the CJPL administration and the Yalong River Hydropower Development Company Ltd for indispensable logistics and other supports.

\end{document}